\documentclass[
    prl, twocolumn, superscriptaddress, nofootinbib, amsmath, amssymb,
    aps, floatfix, preprintnumbers, 10pt
]{revtex4-2}
\usepackage{mathrsfs}
\usepackage{lipsum}
\usepackage{tabularx}
\usepackage{setspace}
\usepackage[utf8]{inputenc}
\usepackage[dvips]{graphicx}
\usepackage[dvipsnames]{xcolor}
\usepackage{float}
\definecolor{CiteBlue}{RGB}{45,52,151}
\usepackage[
    colorlinks=true,
    linkcolor=CiteBlue,
    urlcolor=CiteBlue,
    citecolor=CiteBlue
]{hyperref}
\usepackage[all]{hypcap}
\usepackage{times}
\usepackage{bm}
\usepackage{xr}
\usepackage{verbatim}
\usepackage{booktabs}
\usepackage{boldline}
\usepackage{threeparttable}
\usepackage{nicematrix}
\usepackage{multirow}
\usepackage{amsmath}
\usepackage[en-GB]{datetime2}
\usepackage{upgreek}
\usepackage{comment}

\newcommand{\ee}{\mathrm{e}}
\newcommand{\dd}{\mathrm{d}}
\newcommand{\ii}{\mathrm{i}}
\newcommand{\kvec}{{\bm k}}

\newcommand{\fref}[1]{Fig.~\ref{#1}}

\newcommand{\eref}[1]{Eq.~(\ref{#1})}

\newcommand{\rref}[1]{Ref.~\cite{#1}}
\newcommand{\rrefs}[1]{Refs.~\cite{#1}}

\newcommand{\MPl}{M_\mathrm{Pl}}
\newcommand{\NRH}[1]{N_\mathrm{RH}}
\newcommand{\tRH}[1]{t_\mathrm{RH}}
\newcommand{\TRH}[1]{T_\mathrm{RH}}
\newcommand{\aRH}[1]{a_\mathrm{RH}}
\newcommand{\HRH}[1]{H_\mathrm{RH}}
\newcommand{\wRH}[1]{w}

\begin{document}
\pagenumbering{arabic}

\title{Nonthermal leptogenesis via cosmological gravitational particle production \\ is tested by inflationary gravitational waves}


\author{Tammi Chowdhury}
\email{chowdh64@myumanitoba.ca}
\affiliation{Department of Physics \& Astronomy, 
University of Manitoba, Winnipeg, Manitoba R3T 2N2, Canada}

\author{Leah Jenks}
\email{ljenks3@jh.edu}
\affiliation{William H. Miller III Department of Physics and Astronomy,
Johns Hopkins University, 3400 N. Charles St., Baltimore, MD 21218, U.S.A.}

\author{Edward W. Kolb}
\email{rocky.kolb@uchicago.edu}
\affiliation{Kavli Institute for Cosmological Physics and Enrico Fermi Institute, The University of Chicago, 5640 South Ellis Avenue, Chicago, IL 60637, U.S.A.}

\author{Andrew~J.~Long}
\email{andrewjlong@rice.edu}
\affiliation{Department of Physics and Astronomy, Rice University, 6100 Main St., Houston, TX 77005, U.S.A.}

\author{Evan McDonough}
\email{e.mcdonough@uwinnipeg.ca}
\affiliation{Department of Physics, University of Winnipeg, Winnipeg MB, R3B 2E9, Canada}

\begin{abstract}
We explore the coincidence of scales between cosmic inflation and right-handed neutrinos in seesaw models. We show that inflation models, which will be tested by next-generation CMB experiments, can produce right-handed neutrinos in sufficient abundance to explain the observed baryon asymmetry of the universe. The model can be tested by gravitational wave signatures from cosmic inflation and particle production. 
\end{abstract}

\maketitle
\setcounter{equation}{0}

\textbf{Context and motivation} ---
The cosmological excess of matter over antimatter, often referred to as the baryon asymmetry of the universe (BAU), is an essential component of the Universe as we know it.  
However, the origin of this excess, i.e., the physics of baryogenesis in the first fractions of a second after the Big Bang, remains entirely unknown.  

It is natural to expect that a class of hypothetical particles called heavy Majorana neutrinos (denoted as $N$) \cite{Drewes:2013gca} may play a role in solving the problem of baryogenesis.  
When they interact with Standard Model (SM) particles, they induce a change in lepton number, such as by converting a neutrino into an antineutrino.  
Furthermore, these interactions may be CP-violating, meaning that particles and antiparticles interact at different rates.  
Both of these features of Majorana neutrinos are necessary ingredients for successful baryogenesis~\cite{Sakharov:1967dj}.  
Moreover, heavy Majorana neutrinos are appealing additions to the Standard Model as they help to explain the tiny values of the light neutrino masses through the seesaw mechanism~\cite{Minkowski:1977sc,Mohapatra:1979ia,Yanagida:1979as,Yanagida:1980xy,Mohapatra:1980yp,Schechter:1980gr}, and they are needed for gauge coupling unification~\cite{Georgi:1974sy}.  

Models of baryogenesis using heavy Majorana neutrinos~\cite{Fukugita:1986hr} can be divided into two classes.  
Thermal leptogenesis~\cite{Barbieri:1999ma,Buchmuller:2004nz,Buchmuller:2005eh,Garbrecht:2018mrp} relies on the hot conditions of the early universe to create the Majorana neutrinos. 
However, thermal leptogenesis faces several notable challenges. 
If the primordial plasma did not achieve a temperature above the Majorana neutrino mass scale, typically $M_N = {\cal O}(10^{13}) \, \mathrm{GeV}$ for seesaw models, then too few $N$'s would be created for successful baryogenesis.  
Additionally, since the BAU is generated when the $N$'s drop out of thermal equilibrium and decay, scatterings among particles in the hot plasma threaten to erase the BAU, and washout avoidance imposes strong restrictions on the model.  
Finally, since the new particles and forces arise at very high scales, thermal leptogenesis faces a challenge of falsifiability.  

In the second class of models, the heavy Majorana neutrinos are produced nonthermally via inflaton decay~\cite{Lazarides:1990huy,Campbell:1992hd,Kumekawa:1994gx,Lazarides:1999dm,Asaka:1999yd,Asaka:1999jb,Jeannerot:2001qu,Asaka:2002zu,Allahverdi:2002gz,Senoguz:2003hc,Hahn-Woernle:2008tsk,Buchmuller:2013dja,Croon:2019dfw,Barman:2021tgt,Ghoshal:2022kqp,Barman:2024ujh,You:2024hit,Han:2024qbw,Delepine:2026knj,Bostan:2026ltp}, nonperturbative preheating~\cite{Giudice:1999fb,Garcia-Bellido:2001dqy,Antusch:2018zvu}, or other interactions~\cite{Cataldi:2024pgt,Ghoshal:2026hev}.  
Nonthermal leptogenesis seems unappealing if additional particles and interactions are required to create the heavy Majorana neutrinos.  
However, the authors of \rrefs{Kobayashi:2010cm,Hashiba:2019mzm,Bernal:2021kaj,Co:2022bgh,Fujikura:2022udt,Barman:2022qgt,Flores:2024lzv} have noted that gravitational interactions alone are sufficient to create the needed population of Majorana neutrinos.  

\begin{figure}[t!]
    \includegraphics[width=\linewidth]{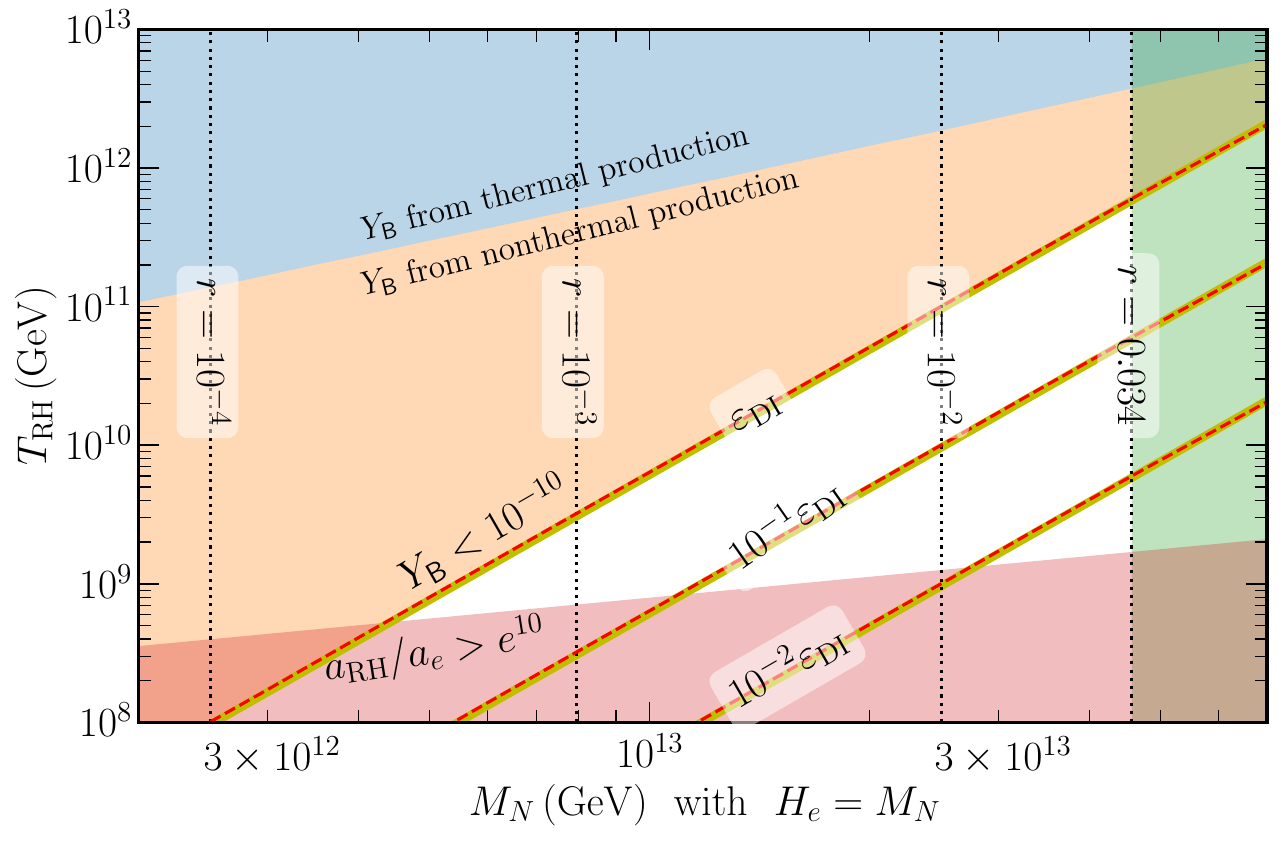}
    \caption{\label{fig:paramspace}
    Parameter space of nonthermal leptogenesis from CGPP. We vary the mass scale of heavy Majorana neutrinos $M_N$, the Hubble scale after inflation $H_e = M_N$, and the plasma temperature after reheating $\TRH{}$.  Within the white triangle the predicted BAU $Y_\mathsf{B}$ matches the observed BAU $Y_\mathsf{B}^\mathrm{obs} \approx 10^{-10}$ for a value of the CP-violation parameter $\varepsilon$ consistent with the Davidson-Ibarra bound $\varepsilon_\mathrm{DI}$.  The green region is incompatible with the observed upper limit on the tensor-to-scalar ratio $r$ \cite{Balkenhol:2025wms}.  The red region is incompatible with a theoretical upper limit on the duration of kination~\cite{Eroncel:2025bcb}.  The orange region is incompatible with baryogenesis in our scenario, since it predicts $Y_\mathsf{B} < Y_B^\mathrm{obs}$ for $\varepsilon < \varepsilon_\mathrm{DI}$.  The blue region is incompatible with our scenario, since thermal production dominates over nonthermal. 
    }
\end{figure}

In the work reported here, we develop a general scenario for nonthermal leptogenesis from cosmological gravitational particle production (CGPP) at the end of inflation \cite{Kolb:2023ydq} followed by a stiff phase of kination dominated by the inflaton's kinetic energy. 
We establish a clear link between successful baryogenesis and the amplitude of primordial gravitational waves, since both are controlled by the energy scale of inflation. 
This link affords our scenario a welcome element of falsifiability. 
\fref{fig:paramspace} shows a slice of the viable parameter space. 
The predicted gravitational wave signal, parametrized by $r$, is within reach of current and next-generation CMB telescopes \cite{LiteBIRD:2022cnt,SimonsObservatory:2025avm}. 

\textbf{Overview of the scenario} --- 
We begin by presenting an overview of our scenario.  
Afterward we elaborate on each of its key elements: (i) a stiff phase (e.g., kination) between inflation and reheating, (ii) nonthermal production of $N$'s via CGPP at the end of inflation, and (iii) lepton-number generation by out-of-equilibrium CP-violating decays.  
Here we derive an analytical expression for the predicted BAU, and later we validate our approximations with a Boltzmann calculation.  

For concreteness we adopt the Type I seesaw model~\cite{Minkowski:1977sc,Mohapatra:1979ia,Yanagida:1979as,Yanagida:1980xy,Mohapatra:1980yp,Schechter:1980gr}, in which the SM is extended by three heavy Majorana neutrinos $N_i$ with masses $M_i$ (for $i = 1, 2, 3$) on the order $M_i = \mathcal{O}(10^{13} \, \mathrm{GeV})$. 
They interact with the Higgs doublet $\Phi_a$ and the lepton doublets $L_{a,i}$ via the Yukawa interaction $\mathscr{L}_\mathrm{int} = - y_{ij} \epsilon_{ab} \bar{L}_{a,i} N_j \Phi_b^\ast  + \mathrm{h.c.}$, where $y_{ij}$ is a 3-by-3 matrix of complex Yukawa couplings.  
After electroweak symmetry breaking $\langle \Phi \rangle = (0, v / \sqrt{2})$ where $v \approx 246 \, \mathrm{GeV}$. 
The seesaw relation gives the light neutrino mass scale $m_\nu = \mathcal{O}(y^2 v^2 / M_N)$ in terms of the heavy neutrino mass scale $M_N$.  

A heavy Majorana neutrino is unstable toward the decay into a lepton $N_j \to L_{a,i} \Phi_b$ or an anti-lepton $N_j \to \bar{L}_{a,i} \bar{\Phi}_b$. 
These rates differ if the interactions violate CP (tied to phases in $y_{ij}$ and loops of heavier $N$'s). 
The dimensionless parameter 
\begin{equation}\label{eq:epsilon_def}
    \varepsilon_j \equiv \frac{
    \Sigma_{ab} \Sigma_i [\mathrm{Br}(N_j\rightarrow L_{a,i} \Phi_b) - \mathrm{Br}(N_j \rightarrow \bar{L}_{a,i} \bar{\Phi}_b)]
    }{
    \Sigma_{ab} \Sigma_i [\mathrm{Br}(N_j\rightarrow L_{a,i} \Phi_b) + \mathrm{Br}(N_j \rightarrow \bar{L}_{a,i} \bar{\Phi}_b)]
    }
\end{equation}
quantifies the CP violation. 
We assume a hierarchical spectrum, $M_{1} \ll M_{2} < M_{3}$, and the BAU arises predominantly from the decays of the lightest neutrino $N_1$, which has \cite{Covi:1996wh} 
\begin{equation}
    \varepsilon_1 = - \frac{3}{16\pi}\frac{1}{(yy^{\dagger})_{11}}\sum_{i=2,3}\text{Im}[(yy^{\dagger})^2_{1i}]\frac{M_1}{M_i} .
\end{equation}
Henceforth we drop the subscript and write $\varepsilon \equiv \varepsilon_1$ and $M_N \equiv M_1$.  
Since $y_{ij}$ also controls the light neutrino mass scale $m_\nu$, the amount of CP violation is constrained by the so-called Davidson-Ibarra bound \cite{Hamaguchi:2001gw,Davidson:2002qv,Buchmuller:2003gz,Hambye:2003rt}: 
\begin{equation}
\label{eq:DIbound}
    |\varepsilon| < \varepsilon_\mathrm{DI} 
    \equiv \frac{3}{8 \pi} \frac{M_{N} m_\nu}{v^2} 
    \approx 9.86 \times 10^{-4} \biggl( \frac{M_N}{10^{13} \, \mathrm{GeV}} \biggr) 
    \;,
\end{equation}
where we set $m_\nu = 0.05 \, \mathrm{eV}$ as the atmospheric mass splitting. 

To realize nonthermal leptogenesis, we assume that a population of $N_1$'s are present at the end of inflation, and we assume that they are sufficiently weakly interacting with the plasma of Standard Model particles so as to not thermalize.  
We demonstrate how this situation naturally arises from CGPP in the following discussion.  
Let $n_N(t)$ denote the number density of $N_1$'s at cosmic time $t$ in a homogeneous and isotropic Friedmann-Robertson-Walker spacetime with scale factor $a(t)$ and Hubble parameter $H(t)$.  
Similarly let $n_\mathsf{L}(t)$ denote the density of lepton number; the left-chiral leptons contribute $n_\mathsf{L} \supset \sum_{i,a} n_{L_{i,a}} - n_{\bar{L}_{i,a}}$. 
After nonthermal $N_1$ production is complete, and assuming negligible thermal interactions, these densities evolve according to \cite{Covi:1996wh}
\begin{subequations}\label{eq:Boltzmann_simple}
\begin{align}
    & \tfrac{\dd}{\dd t} \bigl( a^3 n_N \bigr) = - \Gamma \bigl( a^3 n_N \bigr) \\ 
    & \tfrac{\dd}{\dd t} \bigl( a^3 n_\mathsf{L} \bigr) = + \varepsilon \Gamma \bigl( a^3 n_N \bigr)  
    \;, 
\end{align}
\end{subequations}
where $\Gamma = (y y^\dagger)_{11} M_N / 8\pi$ denotes the total decay rate of $N_1$'s at tree level.  
All of the $N_1$'s have decayed by $t \gg \Gamma^{-1}$. 
Each decay produces $\varepsilon$ units of lepton number on average.  
Eventually a fraction of the lepton number is converted into baryon number through the electroweak sphaleron, which introduces a factor of $-28/79$ \cite{Harvey:1990qw}.  
The resultant comoving number density of baryon number at time $t$ is 
\begin{align}
    a^3(t) n_\mathsf{B}(t) = - \frac{28}{79} \, \varepsilon \, \bigl[ a^3 n_N \bigr]_\mathrm{init}
    \;.
\end{align}
In studies of baryogenesis, it is customary to quantify the BAU using the baryon-to-entropy ratio $Y_\mathsf{B} = n_\mathsf{B}(t) / s(t)$ where $s(t) = \tfrac{2\pi^2}{45} g_\ast(t) T(t)^3$ is the entropy density of a plasma with temperature $T$ and with $g_\ast$ effective relativistic degrees of freedom.  
We use subscript $\mathrm{RH}$ to denote quantities evaluated at reheating, i.e., the beginning of radiation domination. 
We assume that the plasma does not receive any entropy injections during the radiation-dominated epoch ($t \geq \tRH{}$), which implies $a^3(t) s(t) = a^3(\tRH{}) s(\tRH{})$. 
We further assume that the cosmological medium has an effective equation of state $\wRH{}$ between the end of inflation ($t = t_e$) and the start of radiation domination ($t = \tRH{}$), which implies $\HRH{} = (\aRH{}/a_e)^{-3 (1+\wRH{})/2} H_e$.  
Using these relations, the predicted BAU is found to be (see also \rref{Flores:2024lzv}) 
\begin{align}\label{eq:YB}
    Y_\mathsf{B} = - \frac{28}{79} \, \varepsilon \, \frac{\TRH{} H_e}{4 \MPl^2} \, \ee^{3 \wRH{} \NRH{}} \, \biggl( \frac{[a^3 n_N]_\mathrm{init}}{a_e^3 H_e^3} \biggr) 
    \;.
\end{align}
Here $\MPl \approx 2.435 \times 10^{18} \, \mathrm{GeV}$ is the reduced Planck mass, and $\NRH{} = \ln(\aRH{}/a_e)$ is the duration of reheating in $\ee$-foldings. 
Note that a stiffer equation of state, i.e. $\wRH{}$ closer to $1$, tends to increase $Y_\mathsf{B} = n_\mathsf{B}/s$.  
This is because $n_\mathsf{B} \propto \ee^{-3 \NRH{}}$ whereas $s \propto \TRH{}^{3}$, and larger $\wRH{}$ implies smaller $\TRH{}$ for the same $\NRH{}$ (or smaller $\NRH{}$ for the same $\TRH{}$). 

\textbf{Stiff phase} --- 
Our scenario requires inflation to be followed by a stiff phase with large equation of state $\wRH{}$.  
Although our paradigm is more general, here we present a concrete model that exhibits the necessary properties.  
We assume that (single-field, slow-roll) inflation is driven by the potential energy of a scalar inflaton field $\phi$, that the inflaton's kinetic energy drives a phase of kination with $\wRH{} \approx 1$ after inflation \cite{Gouttenoire:2021jhk}, and that eventually the inflaton transfers its remaining energy to the SM plasma leading to the standard radiation phase.  
Inflation that exits to kination is a feature of Quintessential Inflation \cite{Peebles:1998qn,Bettoni:2021qfs,deHaro:2021swo}, $k$-inflation \cite{Armendariz-Picon:1999hyi}, and $G$-inflation \cite{Kobayashi:2010cm} among other models.  
Similar scenarios have been proposed in a string theory context \cite{Apers:2024ffe}. 

Near the end of inflation we model the inflaton potential as 
$V = \tfrac{1}{2} m_\phi^2 \phi^2$ for $\phi > 0$ and $V=0$ for $\phi < 0$.  
If the inflaton is initially displaced to $\phi_i \gtrsim M_\mathrm{Pl}$, then this potential gives a phase of inflation that smoothly transitions into kination. 
Additional interactions are required for reheating (see below). 

The stiff phase cannot continue indefinitely.  
The energy density stored in the fluctuations of the relativistic inflaton field redshift as $\rho_{\delta\phi} \sim a^{-4}(t)$, which decreases more slowly than $\rho_\phi(t) \sim a^{-6}(t)$ as the universe expands.  
The authors of \rref{Eroncel:2025bcb} estimate that a post-inflationary phase of kination cannot last for longer than approximately $10$ $\ee$-foldings, since $\rho_\phi$ eventually gives way to $\rho_{\delta\phi}$.  
On \fref{fig:paramspace} we show the theoretically disallowed region (shaded red) where $\aRH{}/a_e > \ee^{10}$.  

We assume that the inflaton has additional interactions with the SM or new physics that allows an order one fraction of its energy to transfer into radiation before $\rho_\phi$ becomes subdominant to $\rho_{\delta\phi}$.  
For example, the interaction $\mathscr{L}_\mathrm{int} = - g^2 \phi^2 \Phi^\dagger \Phi$ causes the rolling inflaton field to scan the Higgs mass, and Higgs particles can be efficiently produced if its mass passes through zero \cite{Fedderke:2014ura,Amin:2018kkg}.  
Such interactions are necessary, because we find that successful baryogenesis requires a higher reheating temperature than gravitational reheating~\cite{Ford:1986sy} alone can provide. 
Additional discussion of reheating at the end of kination is available in \rrefs{Kunimitsu:2012xx,Nakama:2018gll,Gouttenoire:2021jhk}.  

Assuming that the stiff phase is a period of kination with $\wRH{} = 1$, the predicted baryon-to-photon ratio \eqref{eq:YB} becomes 
\begin{align}
\label{eq:YBbenchmak}
    Y_\mathsf{B} & \simeq \bigl( 1 \times
    10^{-10} \bigr) 
    \biggl( \frac{-\varepsilon}{10^{-1}\varepsilon_\mathrm{DI}} \biggr) 
    \biggl( \frac{M_N}{10^{13} \, \mathrm{GeV}} \biggr) 
    \\ & \times 
    \biggl( \frac{H_e}{10^{13} \, \mathrm{GeV}} \biggr)^{\!2} 
    \biggl( \frac{\TRH{}}{10^9 \, \mathrm{GeV}} \biggr)^{\!-1} 
    \biggl( \frac{[a^3 n_N]_\mathrm{init}/ a_e^3 H_e^3}{10^{-3}} \biggr) 
    \;.
    \nonumber
\end{align}
We fix $m_\nu = 0.05 \, \mathrm{eV}$ and $g_\ast = 106.75$.  
For reference the measured baryon density $\Omega_b h^2 = 0.0224 \pm 0.0001$ \cite{Planck:2018vyg} corresponds to $Y_\mathsf{B}^\mathrm{obs} = (0.879 \pm 0.004) \times 10^{-10}$. 
For these fiducial parameters, the correct BAU is generated provided that the nonthermal production of $N$'s gives $a^3 n_N$ around $10^{-3} a_e^3 H_e^3$.  

\textbf{Gravitational production of Majorana neutrinos} ---
The phenomenon of cosmological gravitational particle production \cite{Ford:2021syk,Kolb:2023ydq} provides a simple explanation for the heavy Majorana neutrinos that our scenario of nonthermal leptogenesis relies upon.  
CGPP is the process by which massive particles are produced due to the rapid expansion of spacetime \cite{Parker:1969au,Parker:1971pt}, typically during inflation and at the end of inflation. 
The CGPP of spin-1/2 particles \cite{Kuzmin:1998kk,Kuzmin:1999zk,Chung:2011ck,Ema:2019yrd}, like our heavy Majorana neutrinos, is most efficient if their mass is close to the inflationary Hubble scale $M_N \approx H_e$. 
For heavier particles, production is kinematically suppressed; and for lighter particles, production is suppressed by approximate conformal symmetry.  
Given the coincidence of the seesaw scale $M_N \approx 10^{13} \, \mathrm{GeV}$ and the scale of inflation $H_\mathrm{inf} \gtrsim H_e \approx 10^{13} \, \mathrm{GeV}$, CGPP offers a natural explanation for the nonthermal origin of heavy Majorana neutrinos. 

We calculate the spectrum and abundance of gravitationally produced heavy Majorana neutrinos by following the procedures laid out in Ref.~\cite{Kolb:2023ydq}.  
The cosmological expansion is described by the spatially-flat FRW metric with cosmic time $t$, scale factor $a(t)$, and Hubble parameter $H(t)$.  
We assume that the expansion is driven by a real scalar inflaton field $\phi(t)$ that is initially displaced to $\phi(t_i) = \phi_i > 0$ and subsequently evolves on the potential $V(\phi)$.  
Inflation ends when $\dd^2 a / \dd t^2$ changes from positive to negative, and we use subscript $e$ to denote quantities evaluated at the end of inflation. 
Earlier studies of CGPP during the transition from inflation to kination include \rrefs{deHaro:2019oki,Haro:2019umj,Salo:2021vdv,deHaro:2022ukj,Chun:2009yu,Lankinen:2016ile}.  

The heavy Majorana neutrino corresponds to a fermionic spinor field, and its positive- and negative-helicity Fourier modes $U_{\kvec,\pm}(\eta,{\bm x})$ must satisfy the equations of motion~\cite{Kolb:2023ydq}
\begin{equation}
    \ii \frac{\dd}{\dd \eta} \begin{pmatrix} u_{A,k} \\ u_{B,k} \end{pmatrix} = \begin{pmatrix} a M_N & k \\ k & -a M_N \end{pmatrix} \begin{pmatrix} u_{A,k} \\ u_{B,k} \end{pmatrix} \;,
\label{eq:ModeEqN}
\end{equation}
where $k = |\kvec|$ is comoving wavenumber and $\eta$ is conformal time.  
We impose Bunch-Davies initial conditions and solve \eref{eq:ModeEqN} numerically. 
From the solutions we calculate the Bogoliubov coefficients $\beta_{\kvec,\pm}$ and
the comoving number density  
\begin{equation}
    a^3 n_N = \int \! \! \frac{\dd^3 \kvec}{(2\pi)^3} \Bigl( |\beta_{\kvec,+}|^2 + |\beta_{\kvec,-}|^2 \Bigr) \;. 
\end{equation}
We repeat this procedure for several values of $M_N$ while holding fixed $\phi_i = 10 \MPl$ and $m_\phi = 2 \times 10^{13} \, \mathrm{GeV}$, which imply $H_e \approx 1 \times 10^{13} \, \mathrm{GeV}$.  
However, the scaled density $a^3 n_N / a_e^3 H_e^3$ only depends upon the dimensionless ratio $M_N / m_\phi$ (or equivalently $M_N / H_e$).  

\begin{figure}[t!]
    \includegraphics[width=\linewidth]{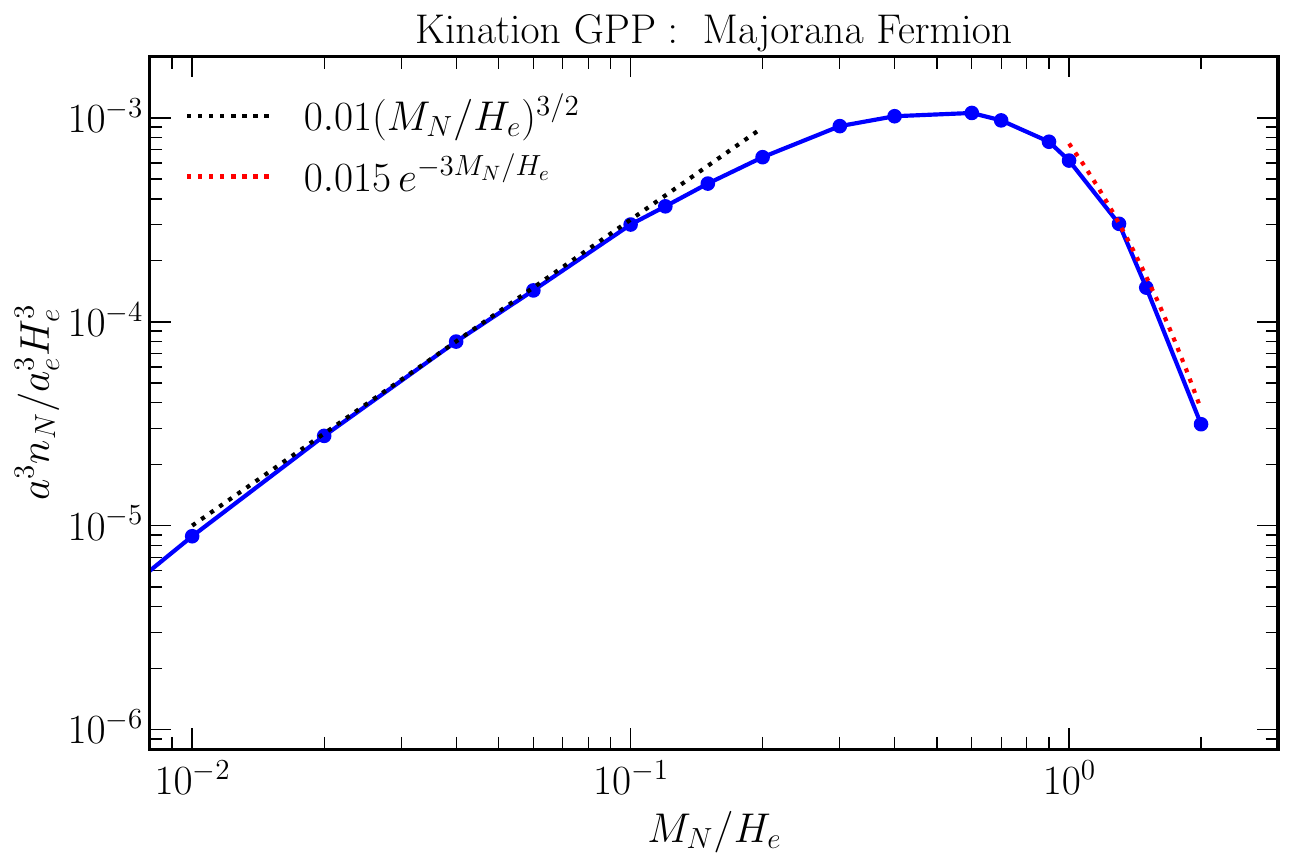}
    \caption{\label{fig:KinGPP}
    Gravitational particle production of Majorana fermions during the transition from an inflationary epoch with a quadratic potential to an era of kination.  The result is that for $10^{-1}\leq M_N/H_e \leq 1$, the value of $a^3n_N/a_e^3H_e^3$ is about $10^{-3}$.  We assume for the fiducial values $M_N=H_e$ and $n_N=6.1\times10^{-4}H_e^3$ at the end of inflation. 
    }
\end{figure}

\fref{fig:KinGPP} shows the predicted abundance of $N$'s that arise from CGPP in this scenario.  
We show the comoving number density $a^3 n_N$ relative to the Hubble scale; for instance, a value $a^3 n_N / a_e^3 H_e^3 = 10^{-3}$ means that there are an average of $10^{-3}$ particles per Hubble volume at the end of inflation. This situation is similar to that of particle production in which inflation is followed by a period of matter domination rather than kination. 
In the former scenario, the spectrum scales as $a^3n_N/(a_eH_e)^3 \propto M_N/H_e$ for $m\lesssim H_e$, and as  $a^3n_N/(a_eH_e)^3 \propto e^{-2M_N/H_e}$ for $M_N\gtrsim H_e$ \cite{Kolb:2023ydq}. 
With an inflationary exit to kination, the low-mass spectrum scales as $a^3n_N/(a_e H_e)^3 \propto (M_N/H_e)^{3/2}$, with a steeper exponential drop: $a^3n_N/(a_e H_e)^3 \propto e^{-3 M_N/H_e}$.
This calculation is consistent with the fiducial parameters used for the estimate in \eref{eq:YBbenchmak}, which gave $Y_\mathsf{B} \sim 10^{-10} \approx Y_\mathsf{B}^\mathrm{obs}$. 
We find that CGPP naturally accommodates the necessary abundance of $N_1$'s for nonthermal leptogenesis. 

\textbf{Boltzmann analysis} --- 
The expression for $Y_\mathsf{B}$ in \eref{eq:YB} was derived by accounting for $N$ decays, but otherwise neglecting reactions among particles in the plasma that might change the numbers of $N$'s or leptons.  
Here we use a system of Boltzmann equations to study nonthermal leptogenesis while accounting for all relevant reactions.  

The number densities $n_N(t)$ and $n_\mathsf{L}(t)$ evolve according to the coupled set of Boltzmann equations
\begin{subequations}\label{eq:boltzmann}
\begin{align}
    \tfrac{\dd}{\dd t} n_N + 3 H n_N 
    & = - (n_N - n_N^\mathrm{eq}) \bigl[ \Gamma + 2 \, n_{\gamma}^\mathrm{eq} \, \langle \sigma v \rangle_1 \bigr] 
    \label{eq:nN} \\ 
    \tfrac{\dd}{\dd t} n_\mathsf{L} + 3 H n_\mathsf{L} 
    & = \varepsilon \Gamma \bigl( n_N - n_N^\mathrm{eq} \bigr) 
    - n_\mathsf{L} \frac{n_\mathsf{L}^\mathrm{eq}}{n_\ell} \Gamma 
    \label{eq:nL} \\
    & \quad 
    - 2 n_\mathsf{L} n_N \langle \sigma v \rangle_1 
    - 4 n_\mathsf{L} n_\Phi^\mathrm{eq} \langle \sigma v \rangle_2 
    \;.
    \nonumber
\end{align}
\end{subequations}
Here $n_a^\mathrm{eq}(t)$ denotes the equilibrium number density for particles of species $a = N_1, \mathsf{L}, \gamma, \Phi$ at temperature $T(t)$.  
We treat the SM particles as massless, we use $T(t) = \TRH{} (a(t) / \aRH{})^{-1}$ after reheating ($t \geq \TRH{}$), and we use since CGPP produces an energy density $\rho_{\rm SM}={\cal O}(H_e^4)$ in SM particles.
The duration of kination is implicitly determined by specifying $H_e$, $\TRH{}$, and the condition that all of the inflaton's kinetic energy is converted into radiation at reheating $\rho_\phi(\TRH{}) = \rho_\mathrm{rad}(\TRH{})$. 
The collision terms account for: (1) heavy Majorana neutrino decays through $N_1 \to L\Phi$ and $N_1 \to \bar{L}\bar{\Phi}$, (2) inverse decays through $L\Phi \to N_1$ and $\bar{L}\bar{\Phi} \to N_1$, (3) scattering with $\Delta \mathsf{L} = \pm 1$ through $N_1 L \to \Phi^\ast \to f\bar{f}$ and $N_1 \bar{L} \to \Phi \to f \bar{f}$ and $f\bar{f} \to \Phi^\ast \to N_1 L$ and $f \bar{f} \to \Phi \to N_1 \bar{L}$ (where $f$ denotes SM fermion, such as the top quark $t$), (4) scattering with $\Delta \mathsf{L} = \pm 2$ through $L\Phi \to N^\ast \to \bar{L}\bar{\Phi}$ and $\bar{L}\bar{\Phi} \to N^\ast \to L\Phi$.  
To obtain these equations, we have adapted Eqs.~(1)~and~(2) of \rref{Buchmuller:2004nz} and adopted their transport coefficients. 

\eref{eq:boltzmann} reduces to \eref{eq:Boltzmann_simple} when the thermal abundance is small, $n_N^\mathrm{eq} \ll n_N$, and when the scatterings can be neglected. 
Since the scatterings change the system's lepton number, they can wash out the BAU if they come into thermal equilibrium.  
However, for the parameter range of interest, we anticipate that washout will be negligible.  
This is thanks in part to the hierarchy $\TRH{} \ll M_N$, which suppresses inverse decays and $\Delta \mathsf{L} = \pm 2$ scatterings.
For instance, we find that the scattering rate is negligible, in the sense that $\langle \sigma v \rangle_2 n_\Phi^\mathrm{eq} \ll H$ at $t = \TRH{}$, provided that the reheating temperature is not too high $\TRH{} \ll 2 \times 10^{13} \, \mathrm{GeV}$.  
Comparing with \fref{fig:paramspace}, one can see that this condition is satisfied across the white region where our scenario allows for the correct BAU. 

\begin{figure}[t!]
    \includegraphics[width=\linewidth]{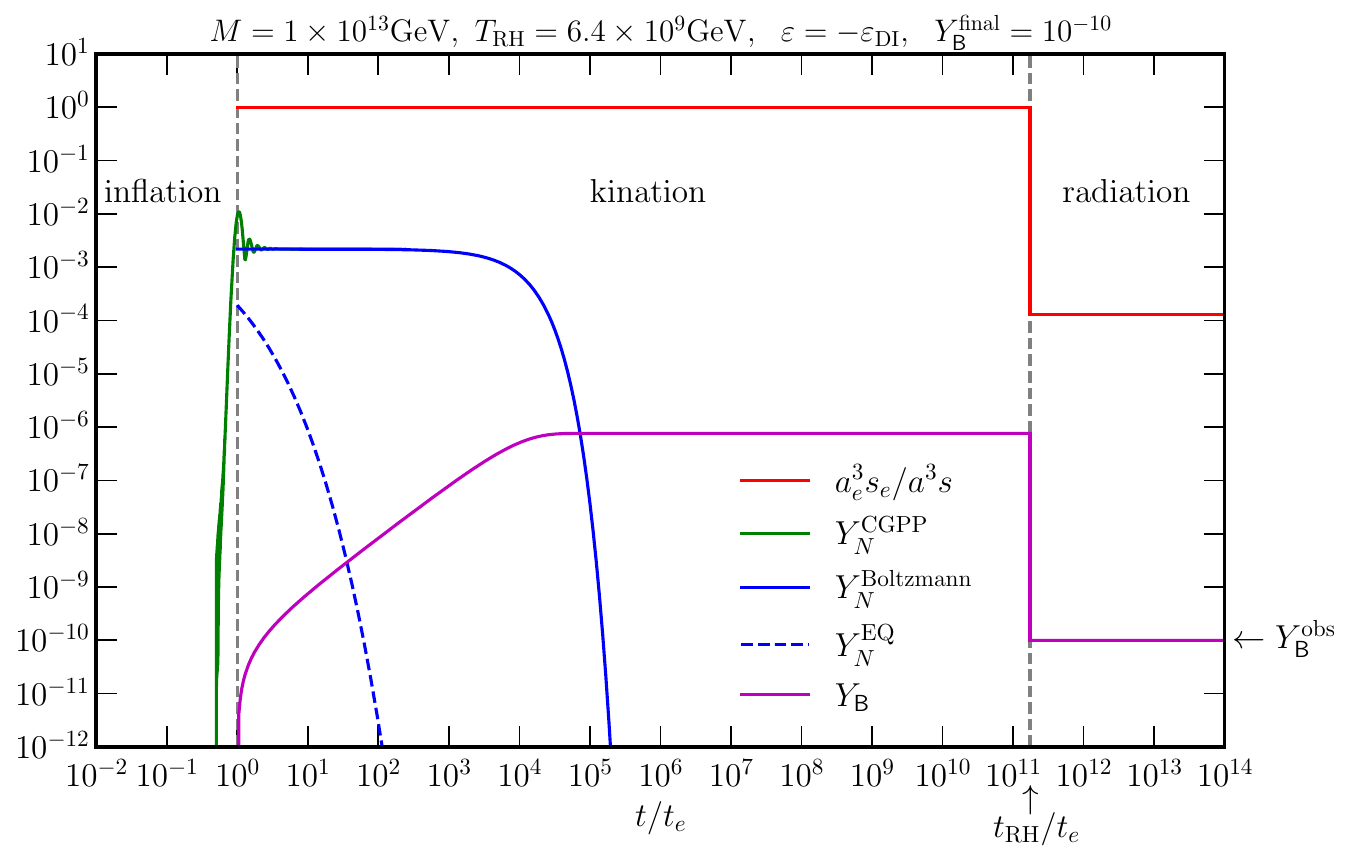}
    \caption{\label{fig:abundance-plot}
    Evolution of abundances.  The density of heavy Majorana neutrinos $Y_N(t) = n_N(t) / s(t)$ (green) grows at the end of inflation ($t \simeq t_e$) due to CGPP and then (blue) decays at the neutrino lifetime ($t = \Gamma^{-1} \approx 10^5 t_e$).  It always remains above the equilibrium density $Y_N^\mathrm{eq}(t) = n_N^\mathrm{eq}(t) / s(t)$ (blue-dashed), signaling that thermal production is negligible.  The decays create lepton number, which is converted into baryon number with $Y_\mathsf{B}(t) = n_\mathsf{B}(t) / s(t)$ (magenta).  At the time of reheating ($t = \TRH{} \approx 10^{11} t_e$), the comoving entropy density $a^3 s(t)$ (red) abruptly increases, and the commensurate reduction in $Y_\mathsf{B}$ brings it into agreement with the observed value $Y_\mathsf{B}^\mathrm{obs} \approx 10^{-10}$. 
    }
\end{figure}

In \fref{fig:abundance-plot} we illustrate the solutions of \eref{eq:boltzmann} for a representative parameter point that lies within the white region of \fref{fig:paramspace}. 
We take $M_N = H_e = 10^{13} \, \mathrm{GeV}$,  $\TRH{} = 6.4 \times 10^{9} \, \mathrm{GeV}$, and $\varepsilon = -\varepsilon_\mathrm{DI} \simeq -9.86 \times 10^{-4}$, 
consistent with the Davidson-Ibarra bound \eqref{eq:DIbound}. 
For the sake of illustration, the figure shows $Y_\mathsf{B} = (-28/79) Y_\mathsf{L}$, which accounts for the eventual conversion of lepton number into baryon number, even though electroweak sphalerons are not yet in equilibrium at these early times.  
After the entropy injection at reheating, we find that the baryon asymmetry settles to $Y_\mathsf{B} \approx 1.0 \times 10^{-10}$.   
The parameters were chosen to give a value close to $Y_\mathsf{B}^\mathrm{obs} \approx 0.879 \times 10^{-10}$, which indicates that baryogenesis is successful.  
Moreover, the result of this numerical Boltzmann analysis is in excellent agreement with our analytic result in \eref{eq:YBbenchmak}, which evaluates to $0.95 \times 10^{-10}$ for these parameters. 
This level of agreement confirms that washout is negligible in the regime $M_N/T \gg 1$ at all times. 

\textbf{Discussion} --- 
\fref{fig:paramspace} shows a slice of our scenario's viable parameter space.  
We vary the heavy Majorana neutrino mass scale $M_N$, the inflationary Hubble scale $H_e = M_N$, and the reheat temperature $\TRH{}$.  
CGPP is most efficient for $H_e \approx M_N$, and we discuss how the parameter space slice shrinks for $H_e \neq M_N$ below.  
We also comment on $\wRH{} \neq 1$. 

The upper-left region of the figure is shaded in blue and labeled ``$Y_\mathsf{B}$ from thermal production.'' 
In this region our scenario of nonthermal leptogenesis is not viable, because thermal production of $N$'s is efficient.
The boundary is defined by equating $a^3 n_N$ from \fref{fig:KinGPP} with $a^3 n_N^\mathrm{eq}$ from \eref{eq:boltzmann} evaluated at $t=\TRH{}$.  

The middle region of the figure is shaded in orange and labeled ``$Y_\mathsf{B}$ from nonthermal production'' and ``$Y_\mathsf{B} < 10^{-10}$.''  
Here the $N$'s arise mainly from CGPP, but their abundance is insufficient to yield $Y_\mathsf{B} = Y_\mathsf{B}^\mathrm{obs} \approx 10^{-10}$ while also satisfying $\varepsilon \leq \varepsilon_\mathrm{DI}$. 
For such large $\TRH{}$, the kination phase is too short to boost $Y_\mathsf{B} \propto \ee^{3 \wRH{} \NRH{}}$ to the observed value.

The white region shows the viable parameter space.  
The predicted BAU is well-approximated by \eref{eq:YBbenchmak}, and we also calculate it numerically by solving \eref{eq:boltzmann}.  
The BAU achieves $Y_\mathsf{B} = 10^{-10}$ across this region for an appropriately chosen CP-violation parameter $\varepsilon$, which remains compatible with the DI bound.  
The duration of kination $\NRH{}$ ranges from $6$ to $10$ e-folds, and $\varepsilon$ ranges from $\varepsilon_\mathrm{DI}$ to $1.8 \times 10^{-3} \varepsilon_\mathrm{DI}$. 

Across the white region, we calculate the amplitude of inflationary primordial gravitational waves by estimating the tensor-to-scalar ratio as $r \approx (1.6 \times 10^{-3}) (H_e / 10^{13} \, \mathrm{GeV})^2$. 
This estimate assumes $H_\mathrm{inf} \approx H_e$, which is the case in many models, such as hilltop inflation. 
The tensor-to-scalar ratio $r$ ranges from $2 \times 10^{-4}$ to the current CMB upper bound of $0.034$. 
Future CMB telescopes will test much weaker primordial gravitational waves; Simons Observatory expects to probe $r = 1.2 \times 10^{-3}$ (or $r = 7 \times 10^{-4}$ for more optimistic assumptions regarding foregrounds and noise) \cite{SimonsObservatory:2025avm}. 
Roughly 91\%\ (97\%)  of the allowed parameter space, by area in \fref{fig:paramspace}, has $r$ above these values. 
We therefore identify the search for $B$-modes as a promising avenue for testing our scenario.  

Relaxing the assumption $H_e = M_N$ tends to reduce the predicted BAU, since CGPP is less efficient (see \fref{fig:KinGPP}).  
Consequently, the white region grows smaller if $H_e \neq M_N$, and it closes off entirely if $H_e \lesssim 0.08 \, M_N$ or $H_e \gtrsim 2.5 \, M_N$. 
Since our scenario leverages the coincidence between the inflationary Hubble scale and the neutrino seesaw scale, it motivates further study into UV embeddings that naturally realize this relation \cite{Garg:2017tds}. 
Similarly, the white region shrinks if the stiff phase is not kination but rather has a smaller equation of state; it closes off entirely for $\wRH{} \lesssim 0.7$. 

\textbf{Conclusion} ---
In this paper we have explored a scenario for nonthermal leptogenesis via gravitational particle production at the end of inflation. 
In comparison with previous studies \cite{Kobayashi:2010cm,Hashiba:2019mzm,Bernal:2021kaj,Co:2022bgh,Fujikura:2022udt,Barman:2022qgt,Flores:2024lzv}, our work is the first to solve the Boltzmann equations including washout from inverse decays and $\Delta \mathsf{L}=1$ and $\Delta \mathsf{L}=2$ lepton-number violating scatterings, to emphasize the testability via tensor-to-scalar ratio, and to comprehensively study the parameter space of the model accounting for the theoretical upper bound on the duration of kination, the CMB bound on the energy scale of inflation, and the Davidson-Ibarra bound on the CP-violation parameter $\varepsilon$. 

Unlike some other models of nonthermal leptogenesis, our scenario features a welcome element of falsifiability.  
The energy scale of inflation controls both the abundance of heavy Majorana neutrinos and the amplitude of primordial gravitational waves, since they both arise as inflationary quantum fluctuations.  
If future observations push the upper bound on $r$ below $10^{-3}$, it would severely constrain our scenario. 

Our work opens several avenues for future study.  
On the model building side, our phenomenological treatment of the kination epoch could be adapted to reflect a UV framework like string theory that naturally accommodates kination \cite{Apers:2024ffe}.  
Along the same lines, one could revisit our baryogenesis calculation using a realistic model for the end of kination and the transition to radiation domination. 
We focus on Type I seesaw, but our work could be extended to Types II and III seesaw models of leptogenesis~\cite{Ma:1998dx,Ma:1998dn}.  
On the experimental side, one could elaborate the flavor structure of the model so as to draw a closer connection with observations of neutrino flavor oscillations.  
In particular, a model that links the CP-violation needed for leptogenesis (parametrized by $\varepsilon$) with the CP-violation that affects flavor oscillations (parametrized by $\delta_\mathrm{CP}$)~\cite{Moffat:2018smo} would expose the model to the scrutiny of upcoming neutrino oscillation experiments: DUNE and Hyper-Kamiokande.  
Finally, as noted by the authors of \rref{Barman:2022qgt}, since the phase of kination after inflation amplifies the spectrum of inflationary gravitational waves at high frequencies \cite{Nakayama:2008wy,Figueroa:2018twl}, this scenario of LG from CGPP may also be tested by interferometer experiments like LISA, DECIGO, and BBO.  

\textbf{Acknowledgments} --- 
We are grateful to Anish Ghoshal, Yuber Perez-Gonzalez, and Jun'ichi Yokoyama for discussion and comments on the draft.  
This material is based upon work supported (in part: A.J.L.) by the National Science Foundation under Grant No.~PHY-2412797.  
T.C. and E.M. are supported by in a MRS-HQP Pooled Resources grant from the McDonald Canadian Astroparticle Physics Research Institute. 
E.M. is supported in part by a New Investigator Operating Grant from Research Manitoba. 
T.C. and E.M. are supported in part by the Arthur B. McDonald Canadian Astroparticle Physics Institute and by the Natural Sciences and Engineering Research Council of Canada.  
L.J. is supported by the Provost's Postdoctoral Fellowship at Johns Hopkins University.  

\bibliographystyle{JHEP}
\bibliography{ref}

\end{document}